\documentclass[a4paper]{article}
\usepackage{graphicx}

\newcommand{\threejm}[6]{\left(\begin{array}{ccc}#1 & #2 & #3 \\ #4 & #5 & #6 \end{array}\right)}

\begin{document}

\huge

\vspace{3cm}

\begin{center}
Stark effect modeling in the detailed opacity code SCO-RCG
\end{center}

\vspace{0.5cm}

\large

\begin{center}
Jean-Christophe Pain$^{a,}$\footnote{jean-christophe.pain@cea.fr (corresponding author)}, Franck Gilleron$^a$ and Dominique Gilles$^b$
\end{center}

\vspace{0.2cm}

\normalsize

\begin{center}
$^a$ CEA, DAM, DIF, F-91297 Arpajon, France\\
$^b$ CEA, DSM, IRFU, F-91191 Gif-sur-Yvette, France
\end{center}

\vspace{0.5cm}

\begin{center}
{\bf Abstract}
\end{center}

The broadening of lines by Stark effect is an important tool for inferring electron density and temperature in plasmas. Stark-effect calculations often rely on atomic data (transition rates, energy levels,...) not always exhaustive and/or valid for isolated atoms. We present a recent development in the detailed opacity code SCO-RCG for K-shell spectroscopy (hydrogen- and helium-like ions). This approach is adapted from the work of Gilles and Peyrusse. Neglecting non-diagonal terms in dipolar and collision operators, the line profile is expressed as a sum of Voigt functions associated to the Stark components. The formalism relies on the use of parabolic coordinates within SO(4) symmetry. The relativistic fine-structure of Lyman lines is included by diagonalizing the hamiltonian matrix associated to quantum states having the same principal quantum number $n$. The resulting code enables one to investigate plasma environment effects, the impact of the microfield distribution, the decoupling between electron and ion temperatures and the role of satellite lines (such as Li-like $1sn\ell n'\ell' - 1s^2n\ell$, Be-like, etc.). Comparisons with simpler and widely-used semi-empirical models are presented.

\section{Introduction and outline of the method}

In hot dense plasmas encountered for instance in inertial confinement fusion (ICF), the line broadening resulting from Stark effect can be used as a diagnostics of electronic temperature $T_e$, density $n_e$ and ionic temperature $T_i$. This represents a challenging task, since from astrophysical dilute plasmas to ultra-dense nuclear fuel in ICF, the density varies by twenty orders of magnitude. The capability of the detailed opacity code SCO-RCG \cite{Pain15a,Pain15b} was recently extended to K-shell spectroscopy (hydrogen- and helium-like ions), following an approach proposed by Gilles and Peyrusse \cite{Gilles15}. Ions and electrons are treated respectively in the quasi-static and impact approximations and the line profile reads 

\begin{equation}
\phi(\nu)\propto\frac{1}{\pi}\int\mathrm{Re}\left[\mathrm{Tr}\{\hat{d}.\hat{X}^{-1}\}\right]W(F)dF,
\end{equation}

\noindent where $\hat{X}=2i\pi\left(\nu+\nu_1\right)-i\hat{H}(F)/\hbar-\hat{\Lambda}_c$, $\nu_1$ being the frequency of the lower state and $\hat{H}(F)=\hat{H}_0-\hat{d}.F$ the hamiltonian of the ion in the presence of an electric field $F$ following the normalized distribution $W(F)$. $\hat{H}_0$ is the hamiltonian without electric field while $\hat{d}$ and $\hat{\Lambda}_c$ represent respectively the dipole and collision operators. The trace (Tr) runs over the various states of the upper level. If $\Delta \nu_D$ is the Doppler width and $w_k$ the weight of the $k^{th}$ Stark component, neglecting non-diagonal terms in dipolar and collision operators, the line profile can be written as a sum of Voigt ($V$) functions (parametrized as in Ref. \cite{Humlicek79}):  

\begin{equation}
\phi(\nu)=\frac{1}{\sqrt{\pi}}\frac{1}{\Delta\nu_D}\int_0^{\infty}W(F)dF\left[\sum_kw_k(F)V(x_k,y_k)\right]\;\; ; \;\; x_k=\frac{\nu-\nu_0-c_k(F)}{\Delta\nu_D}\;\; ; \;\;  y_k=\frac{\langle k|\hat{\Lambda}_c|k\rangle}{2\pi\Delta\nu_D},
\end{equation}

\noindent where $\nu_0$ is the frequency of the line without external field and

\begin{equation}
c_k(F)=\langle k|-\hat{d}.F|k\rangle\;\;\; ; \;\;\; \hat{\Lambda}_c=\frac{4\pi}{3}n_e\left(\frac{e}{\hbar}\right)^2\hat{d}.\hat{d}\left(\frac{2m}{\pi k_BT_e}\right)^{1/2}\ln\left(\frac{\lambda_{DH}Z}{n^2a_0}\right),
\end{equation}

\noindent $\lambda_{DH}$ being the Debye-H\"uckel length.

\section{Hydrogen-like ions}

Stark effect for hydrogenic ions can be calculated in parabolic coordinates using the basis states $|nqm_{\ell}\rangle$, where $q=n_1-n_2$, $n_1$ and $n_2$ being the so-called parabolic quantum numbers, related by $n_1+n_2+|m_{\ell}|+1=n$, $-\ell\leq m_{\ell}\leq\ell$ being the magnetic orbital quantum number. The perturbation $\hat{d}$ is diagonal in this basis and a 2$^{nd}$-order development gives

\begin{equation}
\langle nqm_{\ell}|-\hat{d}.F|nqm_{\ell}\rangle=\frac{3}{2}\frac{ea_0}{Z}nqF-\frac{1}{16}\frac{e^2a_0^2}{(2Ry)}\left(\frac{n}{Z}\right)^4\left(17n^2-3q^2-9m_{\ell}^2+19\right)F^2.
\end{equation}

\begin{figure}[h]
\begin{center}
\includegraphics[width=11cm]{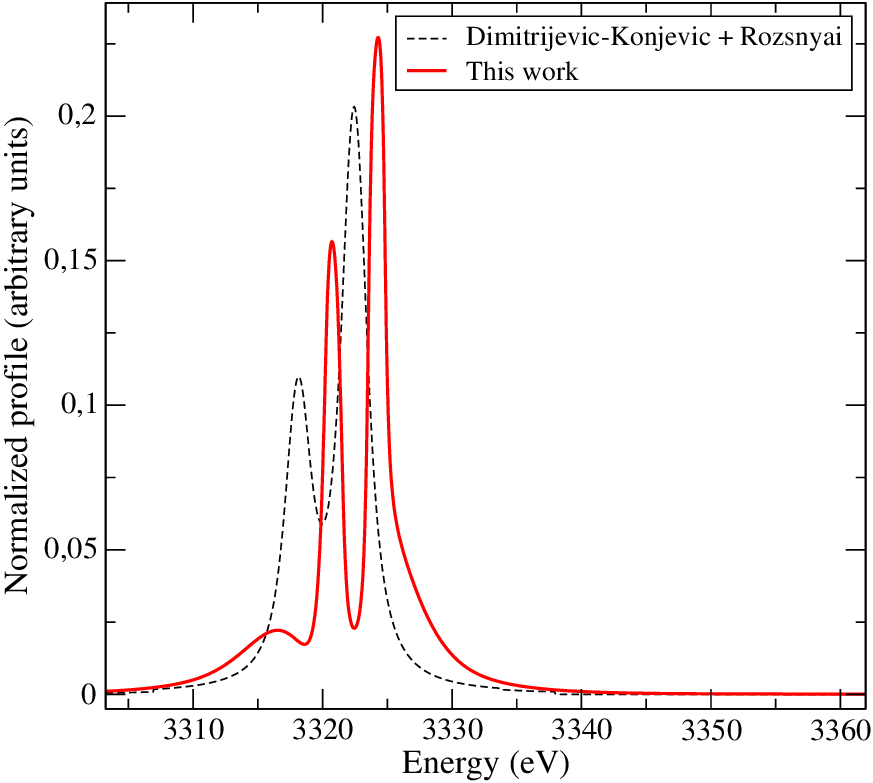}
\vspace{0.5cm}
\caption{\label{fig1} Ly$_{\alpha}$ line for an Ar plasma at $T_e$=$T_i$=700 eV and $\rho$=3.98 g/cm$^3$ (example chosen in Ref. \cite{Mancini13}).}
\end{center}
\end{figure}

\begin{figure}[h]
\begin{center}
\includegraphics[width=11cm]{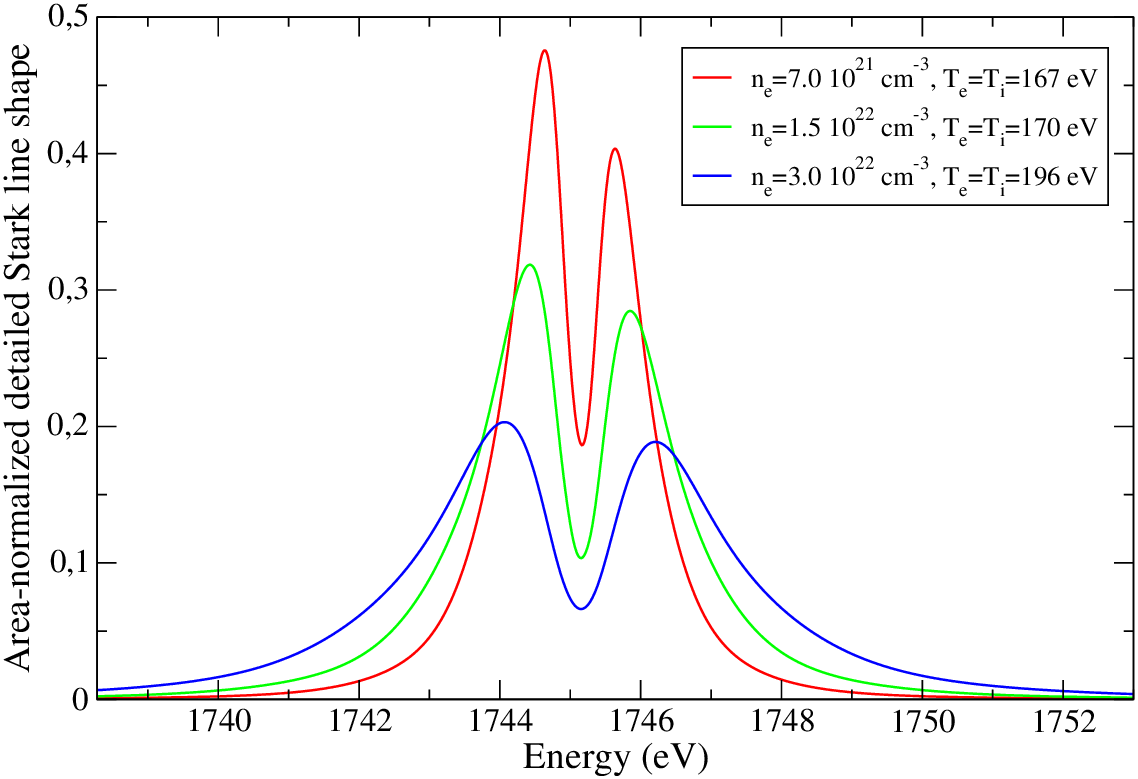}
\vspace{0.5cm}
\caption{\label{fig2} Ly$_{\beta}$ line for a Mg plasma in three different conditions (example chosen in Ref. \cite{Nagayama14}).}
\end{center}
\end{figure}

However, the fine-structure hamiltonian $\hat{H}_0$ is diagonal in the subset of states $|n\ell s j m_j\rangle$. In order to diagonalize the total hamiltonian $\hat{H}$ in such a basis, the Stark matrix element is

\begin{eqnarray}
\langle n\ell sjm_j|-\hat{d}.F|n\ell'sj'm_j\rangle&=&\sum_{m_s=-1/2}^{1/2}\sum_{q=-\left(n-1-|m_{\ell}|\right)}^{n-1-|m_{\ell}|, 2}(-1)^{\ell+\ell'-1+3m_j-m_s-q-n}[\ell,\ell',j,j']^{1/2}\nonumber\\
& &\times\threejm{\ell}{s}{j}{m_{\ell}}{m_s}{-m_j}\threejm{\ell'}{s}{j'}{m_{\ell}}{m_s}{-m_j}\nonumber\\
& &\times\threejm{\frac{n-1}{2}}{\frac{n-1}{2}}{\ell}{\frac{m_{\ell}-q}{2}}{\frac{m_{\ell}+q}{2}}{-m_{\ell}}\threejm{\frac{n-1}{2}}{\frac{n-1}{2}}{\ell'}{\frac{m_{\ell}-q}{2}}{\frac{m_{\ell}+q}{2}}{-m_{\ell}}\nonumber\\
& &\times\langle nqm_{\ell}|-\hat{d}.F|nqm_{\ell}\rangle,
\end{eqnarray}

\noindent with $s=1/2$, $m_{\ell}+m_s=m_j$ and $[x]=2x+1$. Fig. \ref{fig1} displays a comparison between our previous semi-empirical modeling (Refs. \cite{Dimitrijevic80,Rozsnyai77}) and the present work in the case of Ar XVIII Ly$_{\alpha}$ line. Fig. \ref{fig2} shows Mg XII Ly$_{\beta}$ profile in three different conditions.

\section{Helium-like ions}

We consider the transitions $1sn\ell$ $^1P-1s^2$, $n\ge 2$. For $n \geq 5$, the perturbation due to field $F$ is much larger than the separation between terms, the levels are quasi-hydrogenic and He lines are modeled as Ly-like lines (with $Z\rightarrow Z-1$). For $n<5$, singlet-triplet mixing is neglected and two-electron wavefunctions $\Psi(1,2)$ of singlet states are built as \cite{Bethe57}:

\begin{equation}
\Psi(1,2)=\frac{1}{\sqrt{2}}\left[\psi_{100}^Z(1)\psi_{n\ell m_{\ell}}^{Z-1}(2)\pm \psi_{n\ell m_{\ell}}^{Z-1}(1)\psi_{100}^Z(2)\right]\times\chi_{a(+),s(-)}
\end{equation}

\noindent where $\psi_{100}^Z$ is the wavefunction of the fundamental state of an hydrogenic ion of charge Z and $u_{n\ell m_{\ell}}^{Z-1}$ the wavefunction of the excited state $n\ell m_{\ell}$ of an hydrogenic ion of charge $Z-1$. $\chi_{a,s}$ are antisymmetric (resp. symmmetric) spin functions. Such an approximation is valid for highly charged ions when the electron-nucleus interaction overcomes the Coulomb electron-electron repulsion. The hamiltonian $\hat{H}_0-e(z_1+z_2)F$ is diagonalized in the sub-space of states $|1s; n\ell m_{\ell}; S\rangle$ with $S$=0 for singlet states and $S$=1 for triplet states. For He$_{\alpha}$, the resonance line ($1s2p~^1P - 1s^2$) requires the energies of terms $1s2s~^1S$ and $1s2p~^1P$ and the intercombination line ($1s2p~^3P - 1s^2$) the energies of terms $1s2s~^3S$ and $1s2p~^3P$.

\begin{figure}[h]
\begin{center}
\includegraphics[width=11cm]{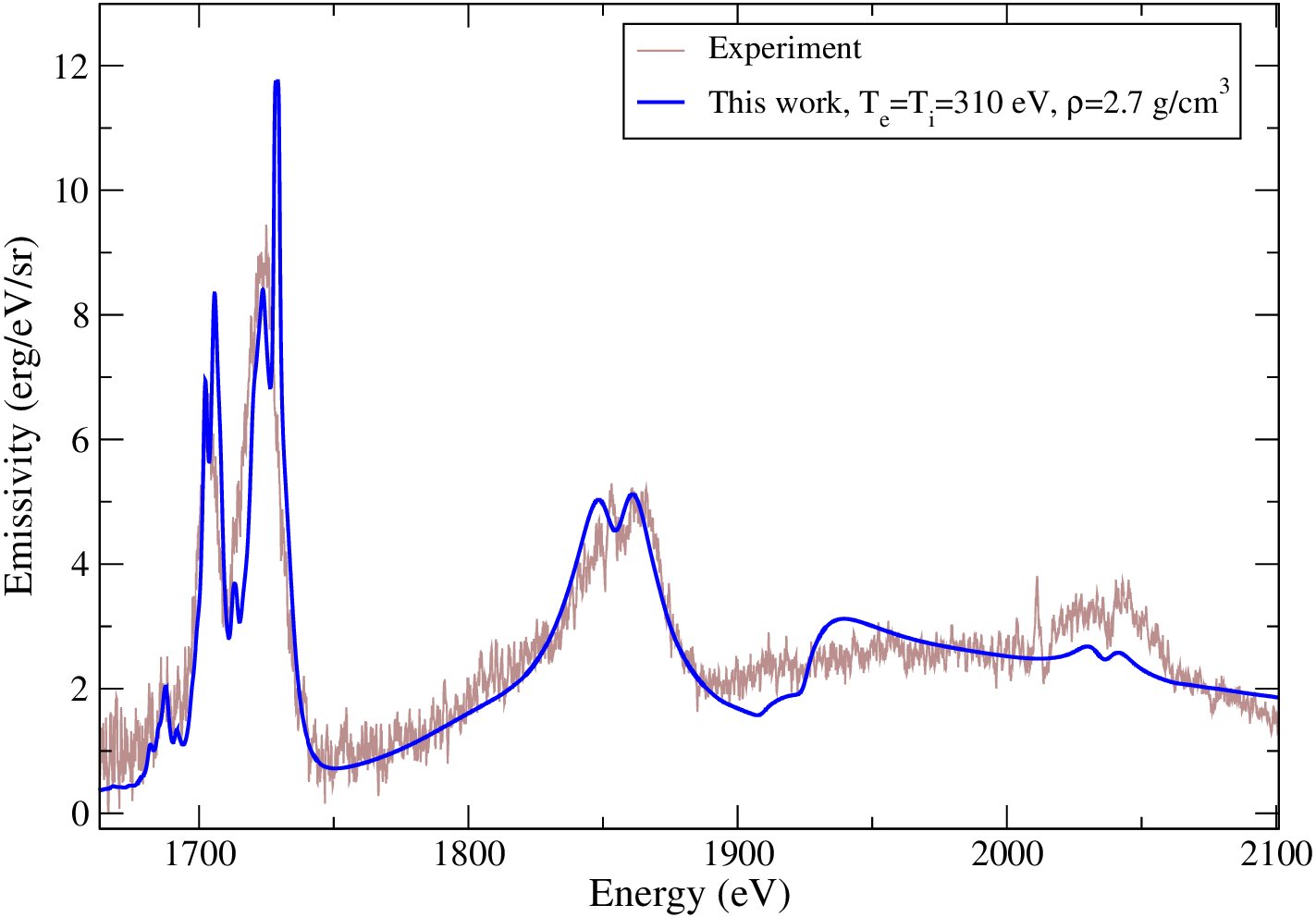}
\vspace{0.5cm}
\caption{\label{fig3} Measured emission of aluminum ``buried layers'' heated by an ultra-short laser \cite{Dervieux15} (emissive volume: 400 $\mu m^2$ $\times$ 0.5 $\mu m$, duration: 3 ps) compared to SCO-RCG prediction.}                      
\end{center}
\end{figure} 

\begin{figure}[h]
\begin{center}
\includegraphics[width=11cm]{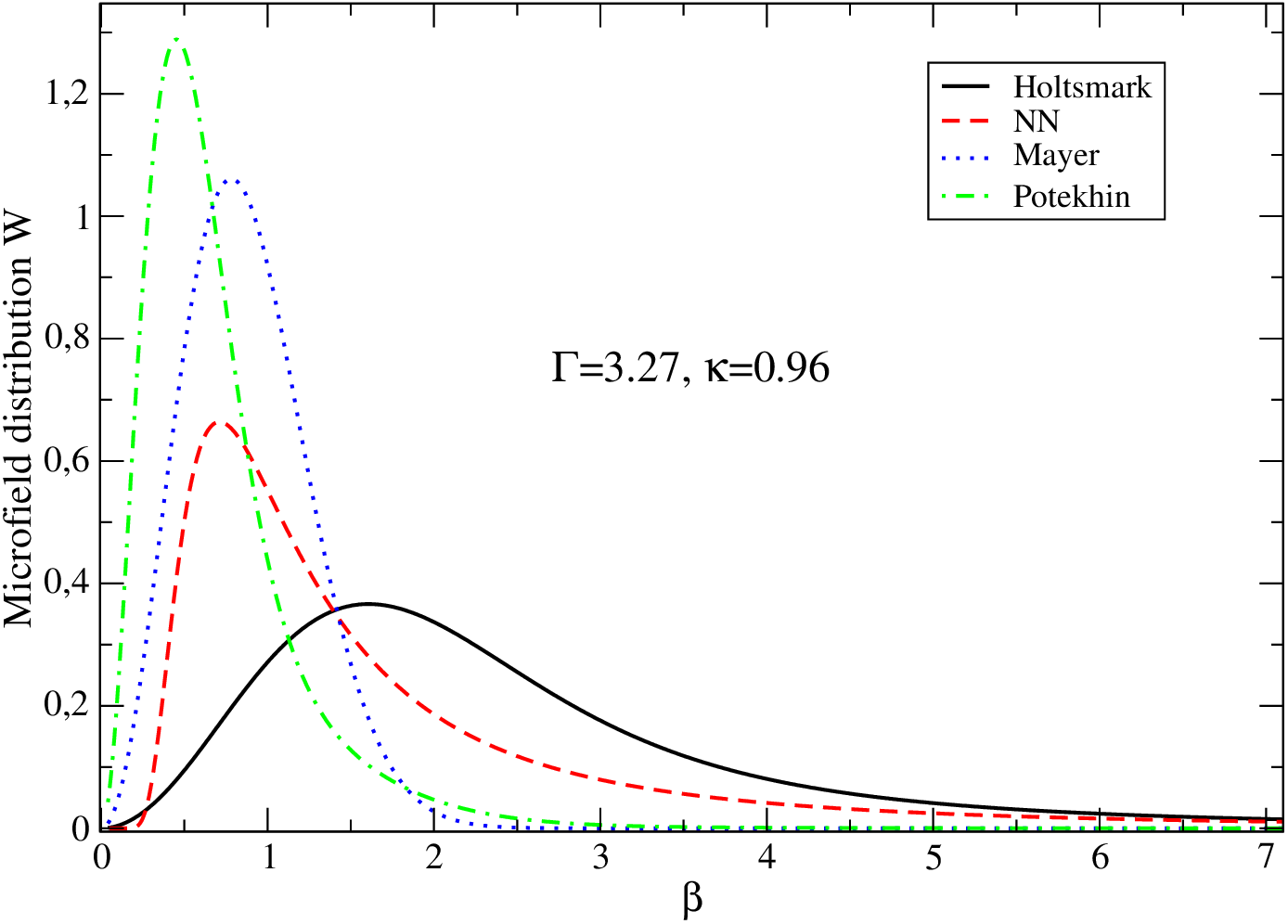}
\vspace{0.5cm}
\caption{\label{fig4} Different microfield distributions for an Al plasma at $T_e$=$T_i$=310 eV and $\rho$=2.7 g/cm$^3$ (units of $\beta=F/F_0$, where $F_0=Z^*e/r_{ws}^2$).}
\end{center}
\end{figure}

\begin{figure}[h]
\begin{center}
\includegraphics[width=11cm]{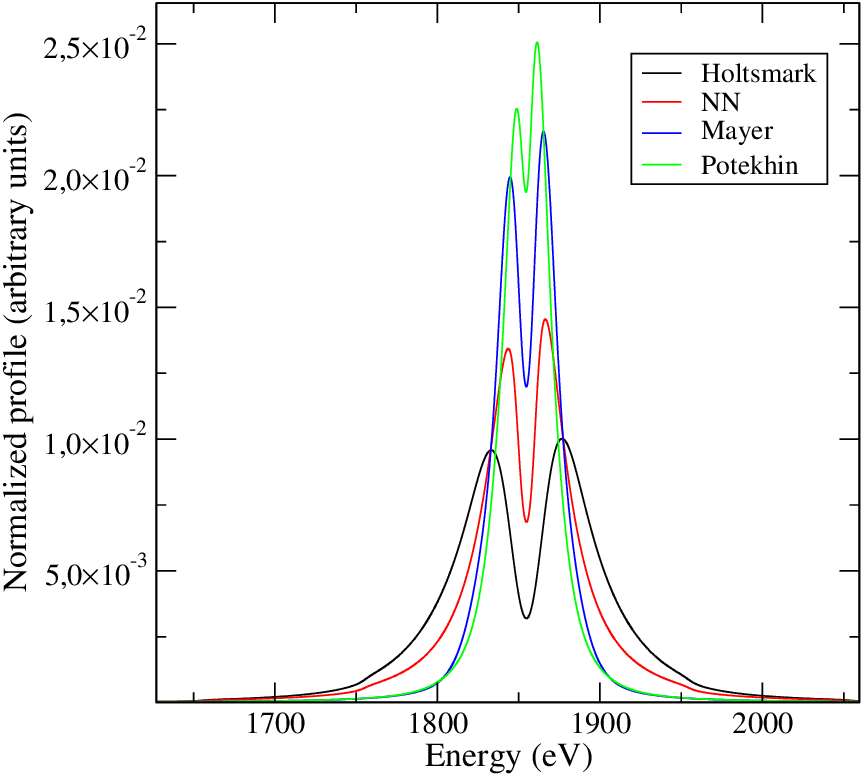}
\vspace{0.7cm}
\caption{\label{fig5} He$_{\beta}$ line for the different microfield distributions of Fig. \ref{fig4}.}
\end{center}
\end{figure}

\section{Interpretation of a ``buried-layer'' experiment on aluminum}

Fig. \ref{fig3} shows our interpretation of the recently measured emission of aluminum micro-targets buried in plastic (``buried layers'') and heated by an ultra-short laser \cite{Dervieux15}. Fig. \ref{fig5} displays the He$_{\beta}$ profile with different models for the microfield distribution function $W(F)$ (see Fig. \ref{fig4}): Holtsmark model, neglecting ionic correlations and electron screening (see Ref. \cite{Potekhin02} and references therein), Mayer and ``nearest neighbour'' (NN) distributions, valid for strongly coupled plasmas, and a combination of APEX (Adjustable Parameter EXponential) method with Monte Carlo simulations proposed by Potekhin et al. \cite{Potekhin02} and parametrized by ionic coupling  $\Gamma=\left(Z^*e\right)^2/\left(r_{ws}k_BT_i\right)$ and electron degeneracy $\kappa=r_{ws}/\lambda_{TF}$ constants, $r_{ws}$ being the Wigner-Seitz radius and $\lambda_{TF}$ the Thomas-Fermi screening length. Table 1 shows the values of the full width at half maximum (FWHM) of Ly$_{\alpha, \beta}$ and He$_{\alpha, \beta}$ lines with the new and previous (Refs. \cite{Dimitrijevic80,Rozsnyai77}) modelings of Stark effect in SCO-RCG.  
                                                                                        
\begin{table*}[t]
\begin{center}
\begin{tabular}{|c|c|c|c|c|}\hline \hline           
Model / FWHM  & Ly$_{\alpha}$ & Ly$_{\beta}$ & He$_{\alpha}$ & He$_{\beta}$  \\ \hline \hline
This work & $1.80$ & $27.03$ & $2.46$ & $32.84$  \\ \hline
Dimitrijevic-Konjevic & $2.19$ & $53.73$ & $2.07$ & $61.92$  \\ \hline \hline
\end{tabular}
\caption{FWHM of Ly$_{\alpha, \beta}$ and He$_{\alpha, \beta}$ lines for an Al plasma at $T_e$=$T_i$=310 eV and $\rho$=2.7 g/cm$^3$.}
\end{center}
\end{table*}
\vspace{0.2cm}

In the future, we plan to investigate the importance of autoionizing states $1s 2\ell 2\ell'$ and $1s2\ell 3\ell'$ of He$_{\beta}$ (in the present work we only took into account $2\ell 2\ell'$) and to include the line $1s3d~^1D_2$ - $1s^2~^1S_0$ induced by the field (mixing states $1s3d~^1D_2$ and $1s3p~^1P_1$) as well as the lines $1s3d~^3D_2$ - $1s^2~^1S_0$ and $1s3s~^3S_1$ - $1s^2~^1S_0$. We also started to study the Stark-Zeeman splitting.

\clearpage

\section{Acknowledgments}

We would like to thank V. Dervieux, B. Loupias and P. Renaudin for providing the experimental spectrum.

\end{document}